\documentclass[10pt,conference]{IEEEtran}

\usepackage{subfigure}
\usepackage{multirow}
\usepackage{graphicx}
\usepackage{amsmath}
\usepackage{color}
\usepackage{epsfig}
\usepackage{amsfonts}
\usepackage{amssymb}
\usepackage{amsthm}
\usepackage[usenames,dvipsnames]{pstricks}
\usepackage{epstopdf}
\usepackage{algorithm}
\usepackage[noend]{algpseudocode}

\newcommand{\argmin}{\mathop{\text{argmin}}}

\newcommand{\vy}{{\bf y}}
\newcommand{\vs}{{\bf s}}
\newcommand{\vn}{{\bf n}}

\newcommand{\mh}{{\bf H}}

\begin{document}

\title{Multidimensional Index Modulation in Wireless 
Communications}
\author{Bharath Shamasundar, Swaroop Jacob, Sandeep Bhat, and 
A. Chockalingam\footnote{This work was supported in part by the 
J. C. Bose National Fellowship, Department of Science and Technology, 
Government of India.} \\
Department of Electrical Communication Engineering \\
Indian Institute of Science, Bangalore 560012, India}
\maketitle

\begin{abstract}
In index modulation schemes, information bits are conveyed through indexing 
of transmission entities such as antennas, subcarriers, times slots, 
precoders, subarrays, and radio frequency (RF) mirrors. Index modulation 
schemes are attractive for their advantages such as good performance, high 
rates, and hardware simplicity. This paper focuses on index modulation 
schemes in which multiple transmission entities, namely, {\em antennas},
{\em time slots}, and {\em RF mirrors}, are indexed {\em simultaneously}. 
Recognizing that such multidimensional index modulation schemes encourage 
sparsity in their transmit signal vectors, we propose efficient signal 
detection schemes that use compressive sensing based reconstruction 
algorithms. Results show that, for a given rate, improved performance 
is achieved when the number of indexed transmission entities is increased. 
We also explore indexing opportunities in {\em load modulation}, which is 
a modulation scheme that offers power efficiency and reduced RF hardware 
complexity advantages in multiantenna systems. Results show that indexing 
space and time in load modulated multiantenna systems can achieve improved 
performance.
\end{abstract}

\vspace{2mm}
{\em {\bfseries keywords:}}
{\em {\footnotesize
Multidimensional index modulation, transmit antennas, transmit RF chains,
time slots, RF mirrors, indexed load modulation, multidimensional 
hypersphere, signal detection, compressive sensing.}} 

\section{Introduction}
\label{sec1}
Conventional modulation schemes convey information bits by sending symbols
from complex modulation alphabets such as quadrature amplitude modulation
(QAM) and phase shift keying (PSK) signal sets. In addition to the bits 
conveyed by complex modulation symbols, the indices of transmission 
entities such as transmit antennas, time slots, subcarriers, subarrays,
etc., which get activated during data transmission can convey additional 
information bits. Modulation schemes in which bits are conveyed through 
such indexing are referred to as index modulation schemes \cite{im5g}. 
Several index modulation schemes and their performance are being reported 
extensively in the recent literature. 

A popular index modulation scheme in the literature is spatial modulation,
where transmit antennas are indexed to convey information bits 
\cite{sm1}-\cite{sm3}. In its basic form, spatial modulation (SM) activates
only one among the available transmit antennas at a time, and which 
antenna gets activated conveys information bit(s). Since only one 
transmit RF chain is adequate for its operation, SM has the RF hardware 
simplicity advantage. A generalized version of SM, referred to as
generalized spatial modulation (GSM) \cite{gsm1}-\cite{gsm3}, 
allows simultaneous activation of more than one antenna at a time,
and which antennas are activated convey information bits. GSM has the
advantage of achieving higher rates and better performance than SM and
spatial multiplexing \cite{gsm2}. SM and GSM employed in a multiuser 
MIMO setting on the uplink have been shown to offer attractive performance 
compared to conventional modulation \cite{mu_sm1}-\cite{mu_gsm}. Similar 
to indexing antennas, subarrays in an antenna array can be indexed. Such 
an indexing scheme, termed as subarray index modulation, can be attractive 
in mmWave communication \cite{saim1},\cite{saim2}. 

Subcarrier index modulation schemes index subcarriers in multicarrier 
systems \cite{sim1}-\cite{sim7}. For example, in an OFDM based subcarrier 
index modulation scheme, not all subcarriers in an OFDM frame carry 
modulation symbols, and which subcarriers among the available subcarriers 
carry symbols convey additional information bits. Equivalently, in the 
time domain, time slots in a frame can be indexed in block transmission 
systems \cite{tim}. That is, some time slots in a frame can be left unused 
by design so that the indices of the used-time slots convey information 
bits. In precoder index modulation \cite{pim1},\cite{pim2}, the transmitter 
is provided with a set of pseudo-random precoder matrices, and, at a time, 
one among them is chosen and used. The index of the precoder used conveys 
information bits. 

The use of parasitic elements for signaling is getting popular because of
its advantages that include improved performance \cite{am1},\cite{am2}.
Media-based modulation (MBM) is one such scheme \cite{mbm1}-\cite{mbm4}.
MBM uses digitally controlled (ON/OFF) parasitic elements external to
the transmit antenna that act as RF mirrors to create different channel
fade realizations which are used as the channel modulation alphabet, and
uses indexing of these RF mirrors to convey information bits. An advantage
of MBM is that the number of bits conveyed through indexing of RF mirrors
grows linearly with the number of mirrors used. This is in contrast with
SM schemes where the number of index bits grow only logarithmically in
number of antennas. MBM signal vectors have good distance properties,
and this enables MBM to achieve better performance compared to conventional
modulation schemes.

Motivated by the performance gains that can be potentially realized
through the use of index bits, in this paper, we
investigate index modulation schemes in which multiple transmission
entities, namely, antennas, time slots, and RF mirrors, are indexed 
simultaneously. Specifically, we consider 1) time-indexed
spatial modulation (TI-SM), where time slots and transmit antennas are
indexed, 2) time-indexed media-based modulation (TI-MBM), where time
slots and RF mirrors are indexed, 3) spatial modulation--media-based
modulation (SM-MBM), where transmit antennas and RF mirrors are indexed,
and 4) time-indexed SM-MBM (TI-SM-MBM), where time slots, transmit
antennas, and RF mirrors are indexed simultaneously. We also propose
efficient signal detection schemes that use compressive sensing based
reconstruction algorithms that exploit the sparsity that is inherently
present in the signal vectors of these schemes. It is found that, for 
a given rate, improved performance can be achieved when more transmission 
entities are indexed.

Recently, the concept of {\em load modulation arrays} is getting 
recognized as a promising approach to realizing massive antenna arrays 
with low RF front-end hardware complexity \cite{lm1},\cite{lm2}. 
It eliminates the need for traditional RF upconversion modules 
(superheterodyne or zero-IF) consisting of DACs, mixers, and filters, 
and requires only one transmit power amplifier (PA) for any number of 
transmit antennas. It achieves this by directly varying (`modulating') 
the antenna impedances (load impedances) that control the antenna 
currents.  While the RF complexity advantages of this modulation 
scheme have been articulated well recently, its performance aspects
need more investigations. In this light, we investigate the potential 
role that indexing can play in improving performance. Our results are 
positive in this regard. In particular, we investigate {\em indexed 
load modulation} schemes, where we index space and time. Our performance 
results show that spatial indexing and time indexing in load modulated 
multiantenna systems can achieve improved performance. 

The rest of the paper is organized as  follows. Multidimensional
modulation schemes that index antennas, time slots, and RF mirrors 
are presented in Sec. \ref{sec2}. Compressive sensing 
based algorithms for detection of multidimensional index modulation 
signals are presented in Sec. \ref{sec3}.  Indexed load modulation 
schemes are presented in Sec. \ref{sec4}. Conclusions are presented 
in Sec. \ref{sec5}. 

\section{Multidimensional index modulation schemes}
\label{sec2}
\begin{figure}
\centering
\includegraphics[width=8.5 cm, height= 4 cm]{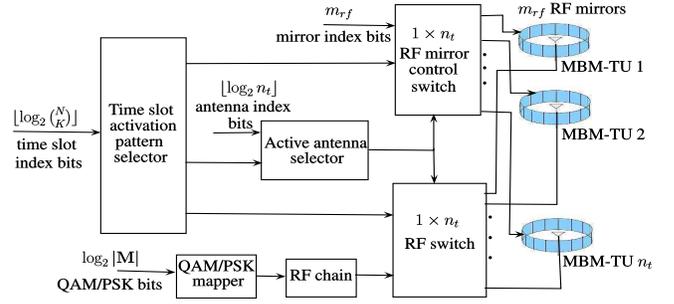}
\caption{Multidimensional index modulation scheme.}
\label{block-diagram}
\end{figure}

In this section, we consider multidimensional index modulation
schemes in which combinations of antennas, time slots, and RF mirrors 
are indexed simultaneously. The considered schemes include 1) TI-SM
scheme, where time slots and transmit antennas are indexed, 2) TI-MBM
scheme, where time slots and RF mirrors are indexed, 3) SM-MBM scheme, 
where transmit antennas and RF mirrors are indexed, and 4) TI-SM-MBM
scheme, where time slots, transmit antennas, and RF mirrors are indexed 
simultaneously. Figure \ref{block-diagram} shows the generalized block 
diagram of the multidimensional index modulation scheme. The notations 
for various system parameters used throughout the paper are listed 
in Table \ref{tab1}.

\begin{table}
\centering
\begin{tabular}{|c|p{0.69\linewidth}|}
\hline
$n_t$ & Number of transmit antennas \\
\hline
$n_r$ & Number of receive antennas \\
\hline
$m_{rf}$ & Number of RF mirrors per transmit antenna \\
\hline
$M  \triangleq 2^{m_{rf}}$ & Number of possible mirror activation patterns (MAP)\\
\hline
$n_L$ & Number of load modulation transmit units (LM-TU) \\
\hline
$n_K$ & Number of active LM-TUs \\
\hline
$n_M$ & Number of vectors in the load modulation alphabet \\
\hline
$N$ & Length of the data part of a frame in no. of time slots \\
\hline
$K$ & Number of active time slots per frame \\
\hline
$L$ & Number of multipaths \\
\hline
$\mathbb{M}$ & Conventional modulation alphabet \\
\hline
$\mathbb{T}$ & Set of all valid time-slot activation patterns (TAP) \\
\hline
$\mathbb{L}$ & Set of all valid LM-TU activation patterns (LAP) \\
\hline
$\mathbf{t^x}$ & TAP corresponding to signal vector $\mathbf{x}$ \\
\hline
${\mathbf q}^{\mathbf{x}}$ & LAP corresponding to signal vector $\mathbf{x}$\\
\hline
$\mathbb{S}_{{\scriptsize{\mbox{sm}}}}$ & Spatial modulation (SM) signal set \\
\hline
$\mathbb{S}_{{\scriptsize{\mbox{mbm}}}}$ & Media-based modulation (MBM) signal set \\
\hline
$\mathbb{S}_{{\scriptsize{\mbox{ti-sm}}}}$ & Time-indexed SM (TI-SM) signal set \\
\hline
$\mathbb{S}_{{\scriptsize{\mbox{ti-mbm}}}}$ & Time-indexed MBM (TI-MBM) signal set \\
\hline
$\mathbb{S}_{{\scriptsize{\mbox{sm-mbm}}}}$ & SM-MBM signal set \\
\hline
$\mathbb{S}_{{\scriptsize{\mbox{ti-sm-mbm}}}}$ & TI-SM-MBM signal set \\
\hline
$\mathbb{S}_{\scriptsize{\mbox{lm}}}$ & Load modulation (LM) alphabet \\
\hline
$\mathbb{S}_{\scriptsize{\mbox{si-lm}}}$ & Spatially-indexed LM (SI-LM) alphabet \\
\hline
$\mathbb{S}_{\scriptsize{\mbox{ti-lm}}}$ & Time-indexed LM (TI-LM) alphabet \\
\hline
$\eta$ & Achieved rate in bits per channel use (bpcu) \\
\hline
\end{tabular} 
\vspace{2mm}
\caption{}
\label{tab1}
\vspace{-8mm}
\end{table}

\subsection{Time-indexed spatial modulation (TI-SM)}
\label{sec2a}
In TI-SM, indexing is done across time and space (i.e., across 
time slots and antennas) \cite{stim}. The TI-SM scheme has $n_t$ 
transmit antennas and one transmit RF chain. Figure \ref{block-diagram}
specializes to TI-SM if RF mirrors and the RF mirror control switch are 
removed. Information bits are conveyed through time-slot indexing, 
antenna indexing, and QAM/PSK symbols. The channel between a 
transmit-receive antenna pair is assumed to be frequency-selective 
with $L$ multipaths. Time-slot and antenna indexing are done as follows.

\subsubsection{Time-slot indexing} 
Time is divided into frames. Each frame  consists of $N+L-1$ time-slots, 
where $N$ is the length of the data part of the frame in number of 
time slots, and $L-1$ is the number of time slots used for transmitting 
cyclic prefix (CP). Out of $N$ time slots, only $K$ time slots, 
$1 \le K\le N$, are used for data transmission. The choice of which $K$ 
slots among the $N$ slots are selected for transmission conveys 
$\lfloor \log_2 \binom{N}{K} \rfloor$  information bits. These bits are 
called `time index bits' and the selected time slots are called `active 
slots' in the frame. An $N$-length pattern of active/inactive status of 
the slots in a frame is called a `time-slot activation pattern' (TAP). 
There are $\binom{N}{K}$ possible TAPs, of which 
$2^{\lfloor \log_2 \binom{N}{K} \rfloor}$ are used and they form the 
set of valid TAPs.

\noindent {\em Example:} Consider $N=4, K=2$. We have $\binom{N}{K}=6$, 
$2^{\lfloor \log _2 \binom{N}{K} \rfloor}=4$. Out of the six 
possible TAPs, the following four TAPs can be taken as the valid TAPs for
signaling --
$\{[1 \ 0 \ 0 \ 1], [1 \ 0 \ 1 \ 0], [0  \ 1  \ 0 \ 1], [1 \ 1 \ 0 \ 0]\}$,
where `1' indicates an active slot and `0' indicates an inactive slot. 

\subsubsection{Antenna indexing} 
In each active slot, one transmit antenna out of $n_t$ antennas is 
selected  based on $\lfloor \log_2 n_t \rfloor$ bits. These bits are 
called `antenna index bits' and the antenna selected is called the 
`active antenna' in that slot. Note that in a TI-SM frame, the active 
antenna can be different in each of the active slots. Also, none of 
the antennas are active in an inactive slot. Since $K$ out of $N$ 
time slots are active in a frame, $K\lfloor \log_2n_t \rfloor$ 
information bits are conveyed through antenna indexing in one frame. 
Further, a symbol from a conventional modulation alphabet $\mathbb{M}$ 
(say, QAM or PSK) is transmitted from the active antenna in an active 
slot. This conveys $\log_2 |\mathbb{M}|$ bits in each active time slot. 
Hence, $K\log_2 |\mathbb{M}|$  bits are conveyed in a frame by 
conventional modulation symbols. Thus, in each active time slot in a 
frame, an $n_t \times 1$ SM signal vector is transmitted. 
The achieved rate in TI-SM scheme is therefore given by
\begin{eqnarray}
\eta_{{\scriptsize {\mbox{ti-sm}}}} & = & \dfrac{1}{N+L-1} \bigg\{ \Big \lfloor \log_2 \binom{N}{K} \Big \rfloor +  K \big(\big \lfloor \log_2 n_t \big \rfloor \nonumber \\ 
& & \hspace{20mm}+ \log_2 |\mathbb{M}| \big) \bigg\}  \  \mbox{bpcu}.
\end{eqnarray} 
   
\subsubsection{TI-SM signal set} 
As noted earlier, an $n_t\times 1$ SM signal vector is transmitted
in an active time slot of a frame, and nothing gets transmitted in an 
inactive slot. The SM signal set is given by
\begin{align}
& \mathbb{S}_{{\scriptsize{\mbox{sm}}}} = \{\mathbf{s}_{j,l} : j=1,\cdots,n_t, \ l=1,\cdots, |\mathbb{M}| \} \nonumber \\ 
& \hspace{10mm} \mbox{s.t} \ \mathbf{s}_{j,l} = [0 \cdots 0 \hspace{-2mm}\underbrace{s_l}_{\mbox{{\scriptsize $j$th coordinate}}}\hspace{-2mm} 0 \cdots 0]^T,  \ \ s_l \in \mathbb{M}.
\label{ss-sm}
\end{align}   
For example, if 
$n_{t}=4$ and $|{\mathbb M}|=2$ (i.e., BPSK), then the SM signal set 
is given by
\begin{equation}
\hspace{-0mm}
\mathbb{S}_{\scriptsize{\mbox{sm}}}=
\left\{
\begin{bmatrix}
1 \\ 0 \\ 0\\ 0
\end{bmatrix}\hspace{-1.5mm},\hspace{-0.5mm}
\begin{bmatrix}
-1 \\ 0 \\ 0\\ 0
\end{bmatrix}\hspace{-1.5mm},\hspace{-0.5mm}
\begin{bmatrix}
0 \\ 1 \\ 0\\ 0
\end{bmatrix}\hspace{-1.5mm},\hspace{-0.5mm}
\begin{bmatrix}
0 \\ -1 \\ 0\\ 0
\end{bmatrix}\hspace{-1.5mm},\hspace{-0.5mm} 
\begin{bmatrix}
0 \\ 0 \\ 1\\ 0
\end{bmatrix}\hspace{-1.5mm},\hspace{-0.5mm}
\begin{bmatrix}
0 \\ 0 \\ -1\\ 0
\end{bmatrix}\hspace{-1.5mm},\hspace{-0.5mm}
\begin{bmatrix}
0 \\ 0 \\ 0\\ 1
\end{bmatrix}\hspace{-1.5mm},\hspace{-0.5mm}
\begin{bmatrix}
0 \\ 0 \\ 0\\ -1
\end{bmatrix}\hspace{-1.0mm}
\right \}. 
\label{mbm_sigset}
\end{equation}
Let ${\bf x}_1,{\bf x}_2,\cdots,{\bf x}_N$ denote the transmitted signal 
vectors in $N$ slots. The TI-SM signal set then can be written as
\begin{align}
\mathbb{S}_{{\scriptsize{\mbox{ti-sm}}}} = & \ \{\mathbf{x} = [\mathbf{x}_1^T \mathbf{x}_2^T \cdots \mathbf{x}_N^T]^T : \mathbf{x}_i \in \mathbb{S}_{{\scriptsize{\mbox{sm}}}} \cup \mathbf{0}, \nonumber \\
 & \ \|\mathbf{x}\|_0=K  \mbox{ and } \mathbf{t^x} \in \mathbb{T}  \},
\label{ss-ti-sm}
\end{align}
where $\mathbf{0}$ denotes $n_t \times 1$ zero vector,  $\mathbb{T}$ 
denotes the set of all valid TAPs, and $\mathbf{t^x}$ denotes the TAP 
corresponding to the signal vector $\mathbf{x}$. The size of TI-SM signal 
set is
$|\mathbb{S}_{{\scriptsize{\mbox{ti-sm}}}}| = 2^{\lfloor \log_2 \binom{N}{K} \rfloor}(n_t|\mathbb{M}|)^K$.
For example, if $N=4, K=2, n_t=4, |\mathbb{M}|=4$, then 
$|\mathbb{S}_{{\scriptsize{\mbox{ti-sm}}}}|=1024$. 
An $Nn_t \times 1$ TI-SM signal 
vector from $\mathbb{S}_{{\scriptsize{\mbox{ti-sm}}}}$ is transmitted 
over $N$ slots in a frame. 

\subsubsection{TI-SM received signal} 
We assume that the channel remains invariant for one frame duration. 
Let $n_r$ denote the number of receive antennas at the receiver.
Assuming perfect channel knowledge at the receiver, after removing
the CP, the $Nn_r \times 1$  received signal vector can be written as
\begin{equation}
\mathbf{y} = \mathbf{Hx}+\mathbf{n},
\end{equation}
where $\mathbf{n}$ is $Nn_r \times 1$  noise vector with 
$\mathbf{n} \sim \mathcal{CN}(\mathbf{0}, \sigma ^2 \mathbf{I})$ and 
$\mathbf{H}$ is $Nn_r \times Nn_t$ equivalent block circulant matrix 
given by
\begin{equation}
\hspace{-0.00mm}
{\mathbf H}=
\begin{bmatrix}
{\bf H}_0 & {\bf 0} & {\bf 0} & \cdots &{\bf H}_{L-1} & \dots & {\bf H}_1  \\
{\bf H}_1 & {\bf H}_0 & {\bf 0} & \cdots & {\bf 0} & \cdots & {\bf H}_2  \\
\vdots  & & & \vdots  & & \\
{\bf H}_{L-1} & {\bf H}_{L-2} & \cdots & {\bf H}_0 & {\bf 0} & \cdots & {\bf 0}  \\
{\bf 0} & {\bf H}_{L-1} & \cdots & {\bf H}_1 & {\bf H}_0 & \cdots & {\bf 0} \\
\vdots & & & \vdots & & \\
{\bf 0} & {\bf 0} & \cdots & \cdots & \cdots & \cdots & {\bf H}_0
\end{bmatrix}\hspace{-0.5mm},
\label{block_circulant}
\end{equation} 
where $\mathbf{H}_l$ denotes the $n_r\times n_t$ channel matrix 
corresponding to the $l$th multipath, $l=0,\cdots,L-1$. Uniform power
delay profile is assumed such that 
$\mathbf{h}_l^k \sim \mathcal{CN}({\bf 0},\frac{1}{L}\mathbf{I})$, 
where $\mathbf{h}_l^k$ is the $k$th column of $\mathbf{H}_l$. 

\subsection{Time-indexed media-based modulation (TI-MBM)}
\label{sec2b}
In this section, we consider TI-MBM. In TI-MBM, indexing is done across 
time slots and RF mirrors \cite{timbm}. The TI-MBM transmitter consists 
of one transmit antenna (i.e., $n_t=1$) and $m_{rf}$  RF mirrors placed 
near it. The transmit antenna and its $m_{rf}$ RF mirrors are together 
called an `MBM transmit unit' (MBM-TU). Figure \ref{block-diagram} 
specializes to TI-MBM by using only one MBM-TU and removing the RF 
switch. Bits are conveyed through time-slot indexing, RF mirror indexing,
and QAM/PSK symbols. Time-slot indexing is done in the same way as in 
TI-SM. A symbol from  the alphabet ${\mathbb M}$ gets sent in each 
active time slot.

\subsubsection{RF mirror indexing}
The propagation environment near the transmit antenna in each active slot 
is controlled by the ON/OFF status of the $m_{rf}$ RF mirrors. The ON/OFF 
status of the $m_{rf}$ mirrors is controlled by $m_{rf}$ information bits. 
These bits are called the `mirror index bits'. An $m_{rf}$-length pattern 
of ON/OFF status of the mirrors in an active slot is called a `mirror 
activation pattern' (MAP). In an active slot, one of the $2^{m_{rf}}$ MAPs 
is selected using $m_{rf}$ information bits. A mapping is done between the 
combinations of $m_{rf}$ information bits and the MAPs. The mapping between 
the MAPs and information bits is made known a priori to both the transmitter 
and the receiver for encoding and decoding purposes, respectively. 
In each active time slot in a frame, a $2^{m_{rf}} \times 1$ MBM signal 
vector is transmitted. The achieved rate in TI-MBM scheme is therefore 
given by 
\begin{eqnarray}
\eta_{{\scriptsize \mbox{ti-mbm}}} & = & \frac{1}{N+L-1} \bigg\{ \Big \lfloor\log _2\binom{N}{K} \Big \rfloor + K\big(m_{rf} \nonumber \\
& & \hspace{20mm} +\log _2|\mathbb{M}|\big) \bigg\} \ \mbox{bpcu}.
\label{spectral_efficiency}
\end{eqnarray}
It is noted that the conventional MBM without time-slot indexing becomes 
a special case of TI-MBM when $K=N$. 

\subsubsection{TI-MBM signal set}
Define ${\mathbb M}_0\triangleq {\mathbb M}\cup 0$ and 
$M\triangleq 2^{m_{rf}}$. The conventional MBM signal set, denoted by 
$\mathbb{S}_{{\scriptsize{\mbox{mbm}}}}$, is the set of $M\times 1$-sized 
MBM signal vectors, which is given by
\begin{eqnarray}
\hspace{-4mm}
\mathbb{S}_{{\scriptsize{\mbox{mbm}}}} & \hspace{-2mm} = & \hspace{-2mm} \left\{\mathbf{s}_{k,q} \in {\mathbb M}_0^M : k=1,\cdots,M, \ q=1,\cdots,|\mathbb{M}| \right \} \nonumber \\ 
& \hspace{-2mm} & \hspace{-2mm}\ \mbox{s.t } \mathbf{s}_{k,q}= [0 \cdots 0 \hspace{-3mm}\underbrace{s_q}_{\mbox{{\scriptsize $k$th coordinate}}}\hspace{-3mm}0 \cdots 0]^T, \ s_q \in \mathbb{M}, 
\label{ss-mbm}
\end{eqnarray}
where  $k$ is the index of the MAP. The size of the MBM signal set is 
$|\mathbb{S}_{{\scriptsize{\mbox{mbm}}}}|= M|{\mathbb M}|$. For example,
for $m_{rf}=2$ and $|{\mathbb M}|=2$ (i.e., BPSK), the MBM signal set 
is given by
\begin{equation}
\mathbb{S}_{{\scriptsize{\mbox{mbm}}}}=
\left\{
\begin{bmatrix}
1 \\ 0 \\ 0\\ 0
\end{bmatrix}\hspace{-1.5mm},\hspace{-1.0mm}
\begin{bmatrix}
-1 \\ 0 \\ 0\\ 0
\end{bmatrix}\hspace{-1.5mm},\hspace{-1.0mm}
\begin{bmatrix}
0 \\ 1 \\ 0\\ 0
\end{bmatrix}\hspace{-1.5mm},\hspace{-1.0mm}
\begin{bmatrix}
0 \\ -1 \\ 0\\ 0
\end{bmatrix}\hspace{-1.5mm},\hspace{-1.0mm}
\begin{bmatrix}
0 \\ 0 \\ 1\\ 0
\end{bmatrix}\hspace{-1.5mm},\hspace{-1.0mm}
\begin{bmatrix}
0 \\ 0 \\ -1\\ 0
\end{bmatrix}\hspace{-1.5mm},\hspace{-1.0mm}
\begin{bmatrix}
0 \\ 0 \\ 0\\ 1
\end{bmatrix}\hspace{-1.5mm},\hspace{-1.0mm}
\begin{bmatrix}
0 \\ 0 \\ 0\\ -1
\end{bmatrix}\hspace{-1.0mm}
\right \}\hspace{-0.5mm}.
\label{mbm_sigset}
\end{equation}
Let ${\bf x}_1,{\bf x}_2,\cdots,{\bf x}_N$ denote the transmitted signal
vectors in $N$ slots. The TI-MBM signal set then can be written as 
\begin{eqnarray}
{\mathbb S}_{{\scriptsize{\mbox{ti-mbm}}}} & \hspace{-2mm} = & \hspace{-2mm} \left \{\mathbf{x} =[\mathbf{x}_1^T \mathbf{x}_2^T \cdots \mathbf{x}_N^T]^T : 
\mathbf{x}_j \in \mathbb{S}_{{\scriptsize{\mbox{mbm}}}} \cup \mathbf{0}, \right .  \nonumber \\
&\hspace{-2mm} & \left . \hspace{-2mm} \|\mathbf{x}\|_0 = K  \mbox{ and } \mathbf{t}^\mathbf{x} \in \mathbb{T} \right \},
\label{ss-ti-mbm}
\end{eqnarray}
where $\mathbb{T}$ denotes the set of valid TAPs and 
$\mathbf{t}^\mathbf{x}$ denotes the TAP corresponding to $\mathbf{x}$.
The size of the TI-MBM signal set is 
$|{\mathbb S}_{{\scriptsize{\mbox{ti-mbm}}}}|=2^{\lfloor\log_2\binom{N}{K}\rfloor}(M|{\mathbb M}|)^K$. 
An $NM\times 1$ TI-MBM signal vector from 
${\mathbb S}_{{\scriptsize{\mbox{ti-mbm}}}}$ is transmitted over $N$ slots
in a frame.

\subsubsection{TI-MBM received signal}
Let $h^{j}(l,k)$ denote the channel gain from the transmit MBM-TU to 
the $j$th receive antenna on the $l$th multipath for the $k$th MAP, where 
$h^j(l,k) \hspace{-0.5mm}\sim\hspace{-0.5mm} \mathcal{CN}(0,\hspace{-0.5mm}\frac{1}{L})$. 
Assume that the channel remains invariant over one frame duration and 
assume perfect timing and channel knowledge at the receiver. After 
removing the CP, the $Nn_r\times 1$-sized received signal vector 
${\bf y}$ can be written as $\mathbf{y} = \mathbf{H} \mathbf{x}+\mathbf{n}$,
where $\mathbf{n}$ is the noise vector of size $Nn_r \times 1$ with
$\mathbf{n}\sim {\cal{CN}}({\bf 0},\sigma^2\mathbf{I})$, and 
$\mathbf{H}$ is the $Nn_r \times NM$ equivalent block circulant 
matrix of the same form as \eqref{block_circulant}, with  ${\bf H}_l$ 
being the $n_r \times M$ channel matrix corresponding to 
the $l$th multipath.

\subsection{Spatial modulation--media-based modulation (SM-MBM)}
\label{sec2c}
In this section, we consider SM-MBM. In SM-MBM, indexing is done across 
MBM-TUs and RF mirrors. The SM-MBM transmitter has $n_t$ MBM-TUs 
and one RF chain. Figure \ref{block-diagram} specializes to SM-MBM by 
removing the time-slot activation pattern selector, as  indexing in time 
is not involved in SM-MBM. Bits are conveyed through MBM-TU indexing,
RF mirror indexing, and QAM/PSK symbols.  

In SM-MBM, all the $N$ time slots in a frame are used for transmission. 
In each time slot, one out of $n_t$ MBM-TUs is selected based on 
$\big\lfloor \log_2 n_t \big\rfloor$  bits, and an 
$n_t\times 1$ SM signal vector from ${\mathbb S}_{{\scriptsize{\mbox{sm}}}}$
gets transmitted across $n_t$ MBM-TUs. The $m_{rf}$ mirrors near the
active MBM-TU are made ON/OFF depending on the MAP chosen based on  
$m_{rf}$ bits. 
Therefore, in addition to the bits conveyed by the SM vector, the 
different channel fade realizations created by the RF mirrors in the
active MBM-TU also convey bits through the MAP index. Hence, the achieved 
rate in SM-MBM is given by
\begin{equation}
\eta_{{\scriptsize \mbox{sm-mbm}}} = \dfrac{N}{N+L-1} \bigg\{ \big \lfloor\log _2n_t \big \rfloor + m_{rf} + \log _2|\mathbb{M}| \bigg\}. 
\label{sm-mbm-rate}
\end{equation}

\subsubsection{SM-MBM signal set} 
The SM-MBM signal set, denoted by 
$\mathbb{S}_{\scriptsize{\mbox{sm-mbm}}}$, is given by
\begin{eqnarray}
\hspace{-5mm}
\mathbb{S}_{\scriptsize{\mbox{sm-mbm}}} & \hspace{-2mm} = \hspace{-2mm} & \{ \mathbf{z} = [\mathbf{z}_1^T  \mathbf{z}_2^T \cdots  \mathbf{z}_{n_t}^T]^T : \mathbf{z}_j = s_j\mathbf{e}_{l_j}, \nonumber \\
& & \hspace{-10mm} l_j \in \{ 1, \cdots, M \} \mbox{ and } \mathbf{s} = [s_1 s_2 \cdots s_{n_t}]^T \in \mathbb{S}_{\scriptsize{\mbox{sm}}} \}, 
\label{ss-sm-mbm}
\end{eqnarray}
where $s_j \in \mathbb{M} \cup 0$, $l_j$ is the index of the MAP chosen 
on the $j$th MBM-TU, and $\mathbf{e}_{l_j}$ is an $M\times 1$ vector whose 
$l_j$th coordinate is `1' and all other coordinates are zero. Note that the 
length of a SM-MBM signal vector is $n_tM \times 1$, and the size of SM-MBM 
signal set is $|\mathbb{S}_{\scriptsize{\mbox{sm-mbm}}}| = n_tM|\mathbb{M}|$.
The transmit vector in $N$ time slots of a TI-SM frame is an $Nn_tM\times 1$
vector given by $\mathbf{x}=[ {\bf x}_1^T {\bf x}_2^T \cdots {\bf x}_N^T]^T$, 
where $\mathbf{x}_i \in \mathbb{S}_{\scriptsize{\mbox{sm-mbm}}}$. The number 
of possible transmit vectors in a frame is $(n_tM|\mathbb{M}|)^N$.

\subsubsection{SM-MBM received signal} 
Let $h^{(j,i)}(l,k)$ denote the channel gain from from the $i$th MBM-TU to 
the $j$th receive antenna on the $l$th multipath for the $k$th MAP, where 
$h^{(j,i)}(l,k) \sim \mathcal{CN}(0,\frac{1}{L})$. In each slot, an 
SM-MBM signal vector from $\mathbb{S}_{\scriptsize{\mbox{sm-mbm}}}$ is sent. 
At the receiver, after removing the CP, the $Nn_r\times 1$ received signal 
vector $\mathbf{y}$ can be written as $\mathbf{y}=\mathbf{Hx + n}$, where 
$\mathbf{n}$ is the $Nn_r \times 1$ noise vector with 
$\mathbf{n} \sim \mathcal{CN}(0, \sigma^2 \mathbf{I})$, and $\mathbf{H}$ 
is the $Nn_r \times Nn_tM$ equivalent block circulant channel matrix of the 
form \eqref{block_circulant}, with $\mathbf{H}_l$ being the $n_r\times n_tM$ 
channel matrix corresponding to the $l$th multipath.

\subsection{Time-indexed SM-MBM (TI-SM-MBM)}
\label{sec2d}
In this section, we consider TI-SM-MBM, which is a generalized scheme 
of which TI-SM, TI-MBM, and SM-MBM are special cases. In TI-SM-MBM, 
time slots, antennas, and RF mirrors are indexed simultaneously (see
Fig. \ref{block-diagram}). The TI-SM-MBM  scheme has $n_t$ MBM-TUs and 
one transmit RF chain. Time indexing is done by choosing $K$ out of 
$N$ time slots in a frame as active slots, based on 
$\big \lfloor \log_2 \binom{N}{K} \big \rfloor$ bits. 
In an active time slot, one of the $n_t$ MBM-TUs is selected based on 
$\lfloor \log_2n_t \rfloor$ bits and activated. The RF mirrors associated 
with the active MBM-TU are controlled by $m_{rf}$ bits that select a MAP. 
A symbol from $\mathbb{M}$ is transmitted from the active MBM-TU in an 
active time slot. 
Therefore, the achieved rate in TI-SM-MBM scheme is given by
\begin{align}
\eta_{{\scriptsize \mbox{ti-sm-mbm}}} = \dfrac{1}{N+L-1} &\bigg\{ \Big \lfloor \log_2 \binom{N}{K} \Big \rfloor + K\big( \big \lfloor\log _2n_t \big \rfloor  \nonumber
\\& + m_{rf}+\log _2|\mathbb{M}| \big) \bigg\} \ \mbox{bpcu}.
\label{ti-sm-mbm-rate}
\end{align}

\subsubsection{TI-SM-MBM signal set} 
In TI-SM-MBM, an SM-MBM signal vector from 
$\mathbb{S}_{\scriptsize{\mbox{sm-mbm}}}$ in (\ref{ss-sm-mbm})
is transmitted in an active 
time slot and nothing gets transmitted in an inactive time slot. Let 
${\bf x}_1,{\bf x}_2,\cdots,{\bf x}_N$ denote the transmit vectors in 
$N$ slots. The TI-SM-MBM signal set is then given by 
\begin{align}
\mathbb{S}_{\scriptsize{\mbox{ti-sm-mbm}}} = &\{ \mathbf{x} =[\mathbf{x}_1^T \mathbf{x}_2^T \cdots \mathbf{x}_N^T]^T : \mathbf{x}_j \in \mathbb{S}_{\scriptsize{\mbox{sm-mbm}}} \cup \mathbf{0},  \nonumber \\
&\| \mathbf{x} \|_0 = K  \mbox{ and } \mathbf{t^x} \in \mathbb{T} \},
\label{ss-ti-sm-mbm}
\end{align} 
where $\mathbf{0}$ denotes the $n_tM \times 1$ zero vector, $\mathbf{t^x}$ 
is the TAP of $\mathbf{x}$ and $\mathbb{T}$ is the set of all valid TAPs. 
The length of a TI-SM-MBM signal vector is thus $Nn_tM \times 1$. Note that 
there are only $K$ non-zero elements out of $Nn_tM$ elements in a TI-SM-MBM 
signal vector, whose positions are decided by the chosen TAP, MAP, and 
active MBM-TU index. The size of the TI-SM-MBM signal set is 
$|\mathbb{S}_{\scriptsize{\mbox{ti-sm-mbm}}}| = 2^{\lfloor\log_2\binom{N}{K}\rfloor}(n_tM|{\mathbb M}|)^K.$
A signal vector from $\mathbb{S}_{\scriptsize{\mbox{ti-sm-mbm}}}$ is 
transmitted in a frame. 
It can be seen that TI-SM, TI-MBM, and SM-MBM become special cases of 
TI-SM-MBM when $i)$ $K<N$, $n_t>1$, $m_{rf}=0$, $ii)$ $K<N$, $n_t=1$, 
$m_{rf}\geq 1$, and $iii)$ $K=N$, $n_t>1$, $m_{rf}\geq 1$, respectively. 

\subsubsection{TI-SM-MBM received signal}
At the receiver, after removing the CP, the $Nn_r\times 1$ received signal 
vector $\mathbf{y}$ can be written as $\mathbf{y}=\mathbf{Hx + n}$, where 
$\mathbf{x} \in \mathbb{S}_{\scriptsize{\mbox{ti-sm-mbm}}}$ is the 
transmit vector of size $Nn_tM \times 1$, and 
$\mathbf{n}$ and ${\bf H}$ are as defined for SM-MBM.

\subsection{ML detection performance}
\label{sec2e}
\begin{figure}
\centering
\includegraphics[width=9.5 cm, height= 6.25cm]{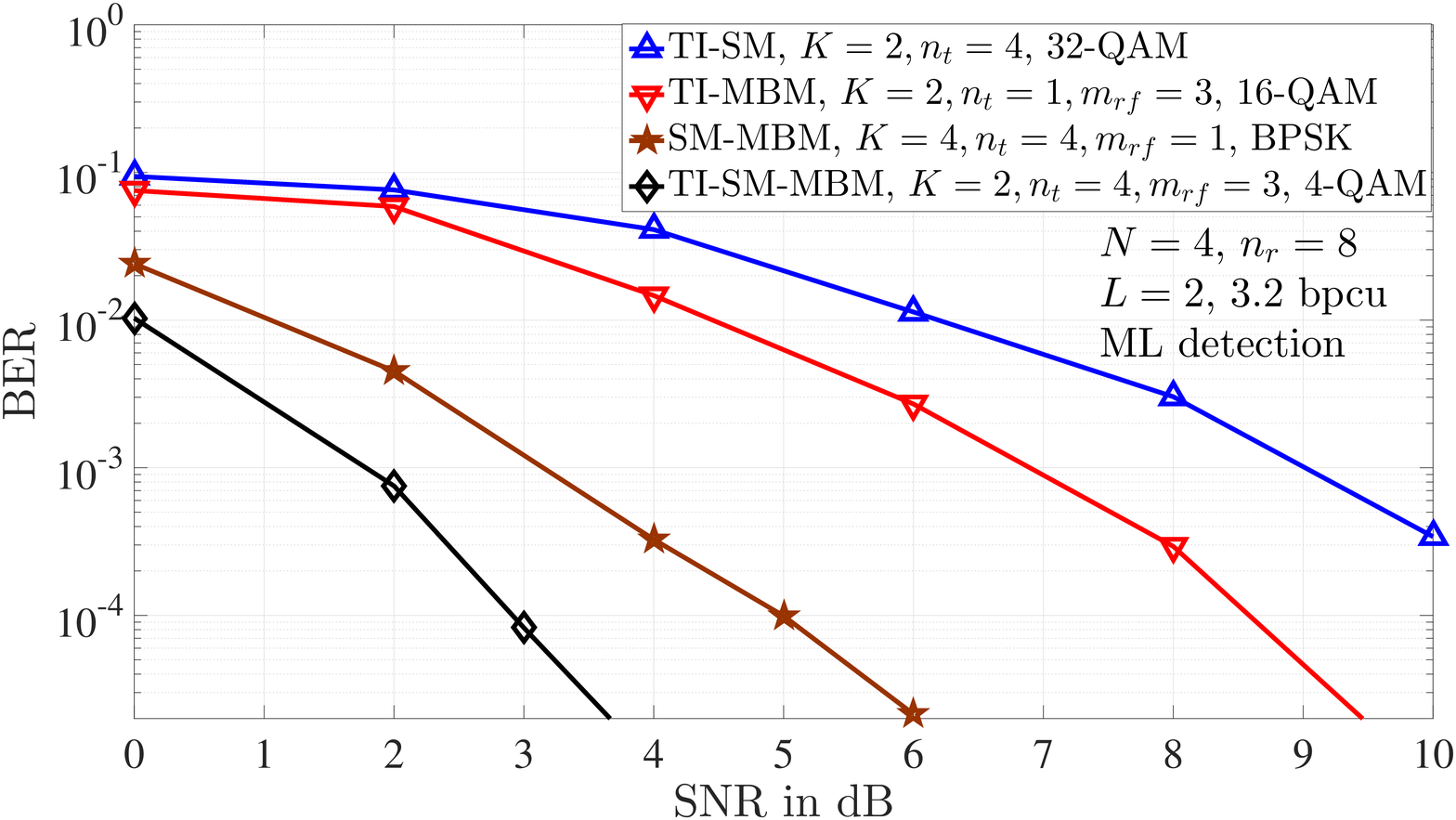}
\vspace{-6mm}
\caption{BER performance of TI-SM, TI-MBM, SM-MBM, and TI-SM-MBM schemes
with 3.2 bpcu under ML detection, $N=4$, $L=2$, $n_r=8$.}
\label{MLD}
\vspace{-4mm}
\end{figure}

In Fig. \ref{MLD}, we present the BER performance of the TI-SM, TI-MBM, 
SM-MBM, and TI-SM-MBM schemes under ML detection for $N=4$, $L=2$, and 
$n_r=8$. Note that in all the four schemes only 
one transmit RF chain is used. All the four schemes are configured such 
that they achieve the same rate of 3.2 bpcu. Also, for the time-indexed
schemes (i.e., for TI-SM, TI-MBM, TI-SM-MBM) $K$ is taken to be 2.
To achieve 3.2 bpcu rate, TI-SM scheme uses $n_t=4$ and 32-QAM, 
TI-MBM uses $n_t=1$, $m_{rf}=3$, and 16-QAM, SM-MBM uses $n_t=4$, 
$m_{rf}=1$, and BPSK, and TI-SM-MBM uses $n_t=4$, $m_{rf}=3$, and 4-QAM. 
The following observations can be made from Fig. \ref{MLD}. 
\begin{itemize}
\item 	First, a comparison between TI-SM and TI-MBM shows that MBM with 
	time indexing can be more attractive compared to SM with time 
	indexing (about 2 dB advantage at $10^{-3}$ BER). This is because 
	to match the bpcu, SM needs more antennas and/or increased QAM-size 
	which can lead to relatively poor performance as observed in the 
	figure. Note that $n_t=4$ and QAM-size is 32 for TI-SM, whereas 
	$n_t=1$ and QAM-size is 16 for TI-MBM. The linear increase in 
	rate as a function of number of mirrors in MBM allows this 
	QAM-size reduction.
\item	Next, SM-MBM is found to outperform both TI-SM and TI-MBM 
	(about 5.8 dB and 3.8 dB advantage, respectively, at $10^{-3}$ 
	BER). This can be attributed to the fact that time-indexed schemes
	incur a rate loss due to inactive time slots, and to compensate
	this loss the QAM-size and/or number of antennas and/or number 
	of RF mirrors have to be increased in order to match the bpcu. 
	This explains why TI-SM and TI-MBM need 32-QAM and 16-QAM, 
	respectively, to achieve 3.2 bpcu. Whereas, SM-MBM achieves 
	the same rate using just BPSK. On the other hand, time indexing 
	offers the advantage of reduced inter-symbol interference (ISI) 
	due to inactive slots. Here, the degrading effect of increased 
	QAM-size dominates the beneficial effect of reduced ISI. 
\item	Finally, TI-SM-MBM in which indexing is done on all the three
	entities (time slots, antennas, RF mirrors) is the most 
	attractive (about 7.2 dB, 5.2 dB, and 1.4 dB advantage at
	$10^{-3}$ BER compared to TI-SM, TI-MBM, and SM-MBM, respectively).
	Note that using time indexing on top of SM-MBM turns out
	to be advantageous compared to SM-MBM without time indexing.
	This is because, here, the beneficial effect of reduced ISI 
	due to time indexing is more compared to the degrading effect 
	of QAM-size increase from BPSK to 4-QAM.
\end{itemize}
The above observations and discussions illustrate that indexing multiple 
entities with a careful choice of system configuration/parameters 
exploiting the underlying tradeoffs involved can be beneficial. Further, 
note that ML detection becomes prohibitively complex for large-dimension 
signals because of the exponential complexity. We exploit the inherent 
sparsity in the indexed transmit vectors to devise low-complexity 
compressive sensing based detection algorithms. 

\section{Compressive sensing based detection}
\label{sec3}
It is noted that the transmit vectors in the multidimensional index 
modulation schemes presented in the previous section are inherently sparse. 
This can be seen from their signal sets defined in \eqref{ss-ti-sm},
\eqref{ss-ti-mbm}, \eqref{ss-sm-mbm}, and \eqref{ss-ti-sm-mbm}. As a 
specific example, consider TI-SM-MBM with $N=16, K=6, n_t=4$, and  
$m_{rf}=4$. While the length of each transmit vector is 
$Nn_tM = 16\times 4 \times 2^4 = 1024$, there are only $K=6$ non-zero 
elements in each vector. This results in a sparsity factor of 
$\frac{K}{Nn_tM}=\frac{6}{1024}$. This sparsity can be exploited for 
efficient signal detection using sparse recovery algorithms.

\subsection{Sparsity-exploiting signal detection}
Several sparse recovery algorithms are known in the literature 
\cite{cs1}-\cite{cs3}. A sparse recovery algorithm seeks 
solution to the following problem:
\begin{equation}
\min\limits_{\mathbf{x}} \| \mathbf{x} \|_0 \mbox{ subject to } \mathbf{b} = \mathbf{Ax + n},
\label{sr}
\end{equation}
where $\mathbf{A} \in \mathbb{C}^{m \times n} $ is the called the sensing 
matrix, $\mathbf{b} \in \mathbb{C}^{m \times 1}$ is the noisy observation 
corresponding to the input vector $\mathbf{x} \in \mathbb{C}^{n \times 1}$, 
and $\mathbf{n} \in \mathbb{C}^{m \times 1}$ is the complex noise vector. 
The sparse nature of the index modulation transmit vectors allows us to 
model the signal detection problem  at the receiver as a sparse recovery 
problem of the form \eqref{sr}. In our detection problem, the sensing 
matrix $\mathbf{A}$ in \eqref{sr} is the channel matrix $\mathbf{H}$, the 
noisy observation $\mathbf{b}$ is the received signal vector $\mathbf{y}$, 
and the goal is to detect the sparse transmit vector ${\bf x}$. We adapt 
greedy sparse recovery algorithms such as orthogonal matching pursuit (OMP) 
\cite{cs1}, compressed sampling matching pursuit (CoSaMP) \cite{cs2}, 
and subspace pursuit (SP) \cite{cs3} for our purpose. For an 
$m\times n$ sensing matrix and a $k$-sparse vector, the reconstruction 
complexity of these algorithms is $\mathcal{O}(kmn)$.
{\bf Algorithm 1} lists the sparsity-exploiting detection algorithm
for TI-SM-MBM.  

\begin{algorithm}[t]
\caption{Sparsity-exploiting detection algorithm}
\begin{algorithmic}[1]
\State Inputs: $\mathbf{y}, \mathbf{H}, K, \mathbb{T} $  	
\State Initialize: $j=0$
\vspace*{3 mm}
\State  \textbf{repeat}
\State \hspace*{0.4 cm} $\mathbf{\hat{x}}_r=\mbox{SR}(\mathbf{y},\mathbf{H},K+j)$ \Comment{Sparse recovery algorithm}
\State \hspace*{0.4 cm} $\mathbf{t}^{\hat{\bf x}_r,(j)}= \mbox{TAP}(\hat{\mathbf{x}}_r)$ \Comment{Extract TAP}
\State \hspace*{0.4 cm} \textbf{if}  $ \| \mathbf{t}^{\hat{\bf x}_r,(j)} \| _0 = K $ and $\mathbf{t}^{\hat{\bf x}_r,(j)} \in \mathbb{T}$ 
\State \hspace*{0.8 cm} \textbf{for} $k= 1 \mbox{ to } N$  
\State  \hspace*{1.2 cm} \vspace*{-5 mm} 
\begin{align*} \hat{\mathbf{x}}^k &= \argmin\limits_{\mathbf{s} \in \mathbb{S}_{\scriptsize{\mbox{sm-mbm}}}} \| \hat{\mathbf{x}}_r^k-\mathbf{s} \|^2,  &&\mbox{\hspace{-8mm}if } t_k^{\hat{\bf x}_r,(j)}=1 \\
& =\mathbf{0}, &&\mbox{\hspace{-8mm}if } t_k^{\hat{\bf x}_r,(j)}=0
\end{align*}
\State \hspace*{0.8 cm} \textbf{end for}
\State \hspace*{0.8 cm}\textbf{break}; \Comment{break repeat loop}
\State \hspace*{0.4 cm}\textbf{else} $j=j+1$ 
\State \hspace*{0.4 cm}\textbf{end if}
\State \textbf{until} $j < (Nn_tM -K )$
\vspace*{3 mm}
\State Output: Estimated TI-SM-MBM signal vector 
$$\hat{\mathbf{x}} = [\hat{\mathbf{x}}^{\scriptsize{1}^T} \hat{\mathbf{x}}^{\scriptsize{2}^T} \cdots \ \hat{\mathbf{x}}^{\scriptsize{N}^T} ]^T$$
\end{algorithmic}
\label{alg1}
\end{algorithm}

In {\bf Algorithm 1}, SR denotes sparse recovery algorithm, which can 
be any one of OMP, CoSaMP, SP. The signal vector reconstructed by SR is 
denoted by  $\mathbf{\hat{x}}_r$. Detecting a TI-SM-MBM signal vector 
involves $i)$ obtaining a valid TAP and $ii)$ detecting the SM-MBM 
signal vector in each active time slot. In a given iteration $j$, the 
estimated TAP, $\mathbf{t}^{\hat{\bf x}_r,(j)}$, is obtained from 
the received vector as follows. A TI-SM-MBM signal vector consists of 
SM-MBM signal vectors in $K$ active time slots and zero vectors in $N-K$ 
inactive time slots. The SM-MBM signal vector in an active time slot 
consists of only one non-zero element. Hence, SR is expected to 
reconstruct $\hat{\mathbf{x}}_r$ with exactly one non-zero element 
in the subvector of each active time slot. This constraint on the 
expected support set is not incorporated in the standard sparse recovery 
algorithms. A standard sparse recovery algorithm can output a vector with 
$K$ non-zero entries in any of the $Nn_tM$ positions of $\hat{\mathbf{x}}_r$. 
In order to extract TAP from $\hat{\mathbf{x}}_r$, the algorithm treats a 
time slot with at least one non-zero entry in the subvector of that time 
slot to be an active time slot. In order to identify the $K$ active time 
slots in the TI-SM-MBM signal vector, SR is used multiple times over a 
range of sparsity values starting from $K$. The TAP vector corresponding
to $\hat{\mathbf{x}}_r$ in the $j$th iteration is obtained such that 
$t_k^{\hat{\bf x}_r,(j)}=1$ if $k$th slot is active and zero otherwise. 
The input sparsity value is incremented by one till a valid TAP 
is obtained. On recovering an $\hat{\mathbf{x}}_r$ with valid TAP, the 
subvector in each active time slot is mapped to the SM-MBM signal vector 
which is nearest in the Euclidean sense (i.e., to the nearest vector in 
$\mathbb{S}_{\scriptsize{\mbox{sm-mbm}}}$ ). This is shown in Step $8$ 
of the algorithm, where $\hat{\mathbf{x}}_r^k$ denotes the 
$n_tM \times 1$-length recovered subvector in the $k$th 
time slot and $\hat{\mathbf{x}}^k$ is the signal vector to which 
$\hat{\mathbf{x}}_r^k$ gets mapped. The detected 
TI-SM-MBM signal vector output from the algorithm is 
$\hat{\mathbf{x}} = [\hat{\mathbf{x}}^{\scriptsize{1}^T} \hat{\mathbf{x}}^{\scriptsize{2}^T} \cdots \hat{\mathbf{x}}^{\scriptsize{N}^T} ]^T$.
The decoding of information bits from $\hat{\mathbf{x}}$ involves 
obtaining time-slot index bits, mirror index bits, antenna index bits, 
and QAM symbol bits. The time-slot index bits are decoded from the 
indices of active time slots in $\hat{\mathbf{x}}$ using combinadics 
based decoding \cite{gsm3}. The mirror index bits and antenna index bits 
are decoded from the detected SM-MBM signal vectors in active time slots. 
The detected SM-MBM signal vector also gives the QAM symbol being 
transmitted, from which QAM bits can be decoded.

We note that {\bf Algorithm 1} can be used for the detection of TI-SM, 
TI-MBM, and SM-MBM, as they are special cases of TI-SM-MBM. The only 
change is in the nearest symbol mapping in Step 8, where $\hat{x}_r^k$ 
has to be mapped to the nearest SM signal vector from 
$\mathbb{S}_{\scriptsize{\mbox{sm}}}$ for TI-SM and to the nearest MBM 
signal vector from $\mathbb{S}_{\scriptsize{\mbox{mbm}}}$ for TI-MBM. 

\subsection{Performance results}
This subsection presents the BER performance of the various index 
modulation schemes with OMP, CoSaMP, and SP based detection. 

\subsubsection{TI-SM-MBM performance with OMP, CoSaMP, SP} 
\begin{figure}
\centering
\includegraphics[width=9.5cm, height=6.25cm]{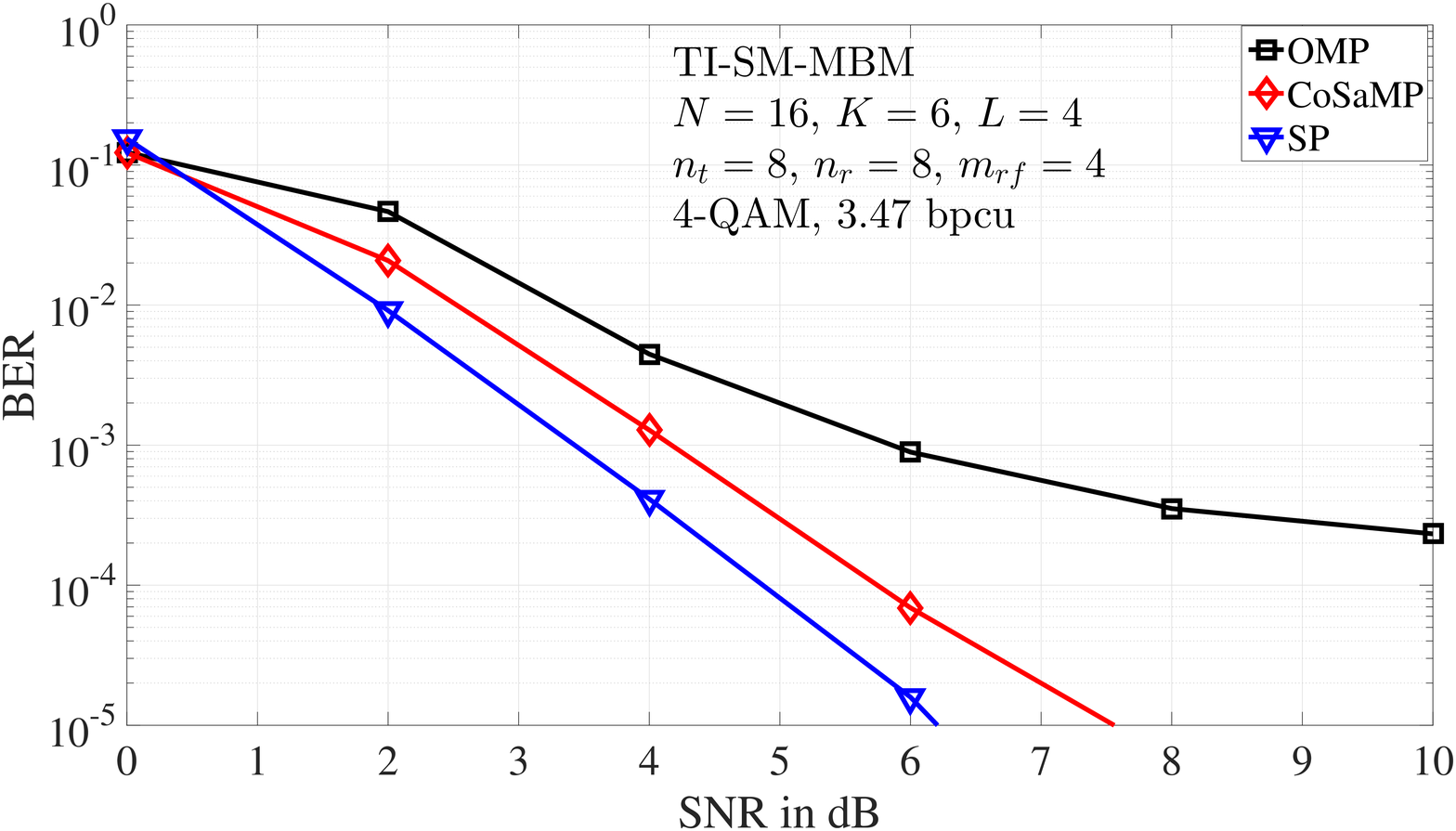}
\vspace{-6mm}
\caption{BER performance of TI-SM-MBM with OMP, CoSaMP, and SP based
detection. $N=16$, $K=6$, $n_t=8$, $n_r=8$, $m_{rf}=4$, 4-QAM, $L=4$, 
3.47 bpcu.}
\label{sparse_algos}
\vspace{-5mm}
\end{figure}

In Fig. \ref{sparse_algos}, we present the BER performance of TI-SM-MBM 
with OMP, CoSaMP, and SP based signal detection. The considered TI-SM-MBM 
scheme uses $N=16, K=6, n_t=8, n_r=8, m_{rf}=4$, and 4-QAM. The number of
multipaths considered is $L=4$, and the achieved rate is 3.47 bpcu. From Fig. 
\ref{sparse_algos}, it can be seen that SP and CoSaMP based detection 
achieve superior performance compared to OMP based detection. While the 
BER floors at 
around $10^{-4}$ with OMP, it falls much below $10^{-4}$ with CoSaMP and 
SP. Between SP and CoSaMP, SP has superior reconstruction performance. 
This can be seen from the SNR advantage of about 1.5 dB at $10^{-5}$ BER 
in favor of SP based detection compared to CoSaMP based detection.

\subsubsection{TI-SM, TI-MBM, SM-MBM, TI-SM-MBM performance with SP based
detection}
\begin{figure}
\centering
\includegraphics[width=9.5 cm,height=6.25cm]{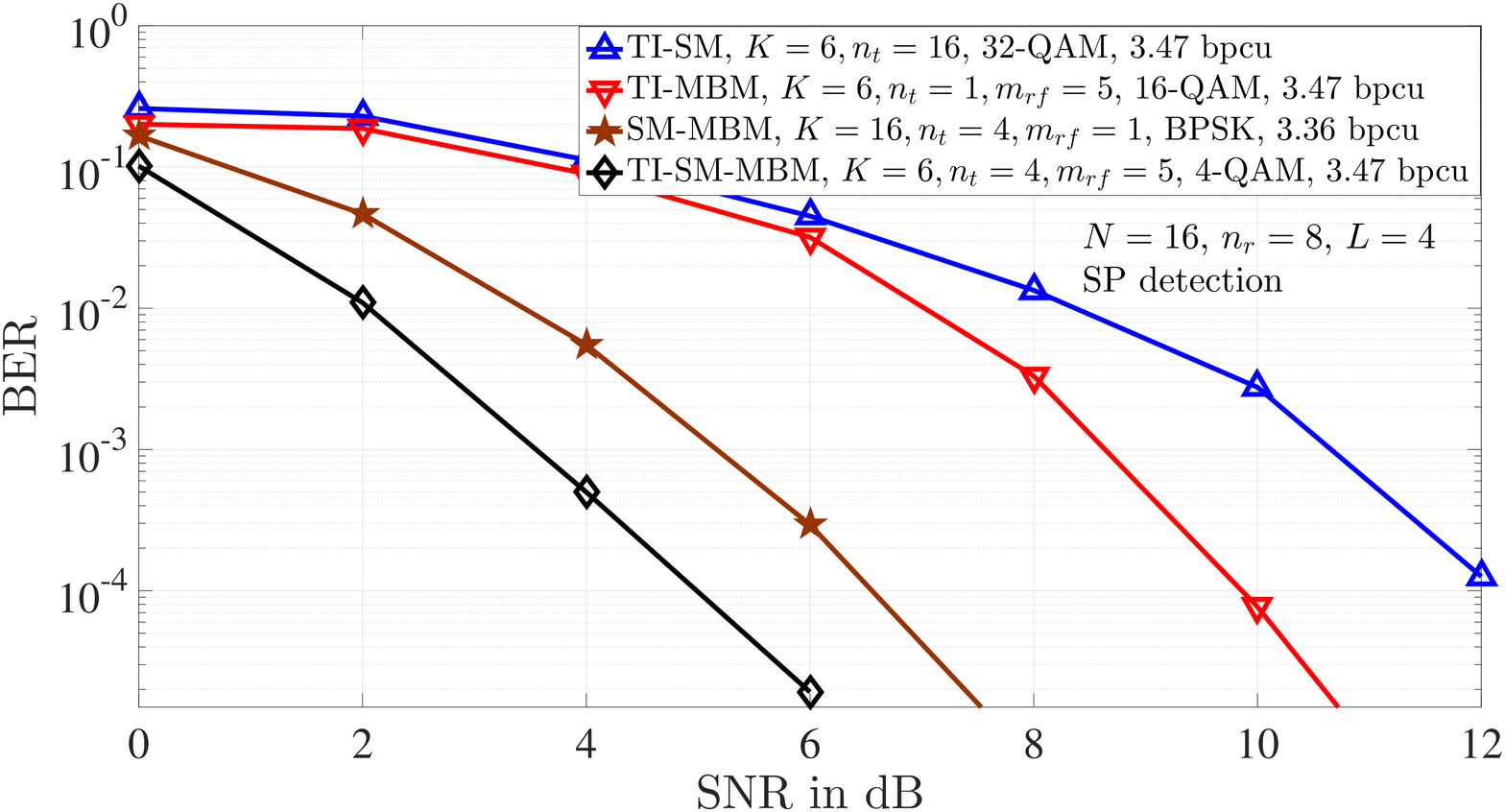}
\vspace{-6mm}
\caption{BER performance comparison of TI-SM, TI-MBM, SM-MBM, and 
TI-SM-MBM schemes with SP based detection.}
\label{sp_detection_mi}
\vspace{-5mm}
\end{figure}

In Fig. \ref{sp_detection_mi}, we present the BER performance of TI-SM, 
TI-MBM, SM-MBM and TI-SM-MBM with SP based detection. All the four schemes
use $N=16$ and $n_r=8$. The number of multipaths is $L=4$. The SM-MBM scheme 
achieves 3.36 bpcu using $n_t=4, m_{rf}=1$, and BPSK. The TI-SM, TI-MBM, 
and SM-MBM schemes achieve 3.47 bpcu using the following configurations:
$i)$ TI-SM: $K=6, n_t=16$, and 32-QAM, $ii)$ TI-MBM: $K=6,n_t=1, m_{rf}=5$, 
and 16-QAM, and $iii)$ TI-SM-MBM: $K=6, n_t=4,m_{rf}=5$, and 4-QAM.
It can be seen from Fig. \ref{sp_detection_mi} that SM-MBM and TI-SM-MBM 
schemes have superior performance compared to TI-SM and TI-MBM schemes. 
Also, TI-SM-MBM performs the best among all the considered schemes. We 
can see that this comparative performance behavior of the considered 
schemes for large-dimension systems ($N=16, K=6$, length of the signal 
vectors: $2^8$ to $2^{11}$) with SP based detection is similar in trend 
compared to what was observed for small-dimension systems ($N=4,K=2$, 
length of the signal vectors: $2^4$ to $2^7$) with ML detection in 
Fig. \ref{MLD}. The reasoning we presented in Sec. \ref{sec2e} for the 
relative performance in Fig. \ref{MLD} applies here as well. 

\subsubsection{Effect of the number of receive antennas, $n_r$}
\begin{figure}
\centering
\includegraphics[width=9.5cm,height=6.25cm]{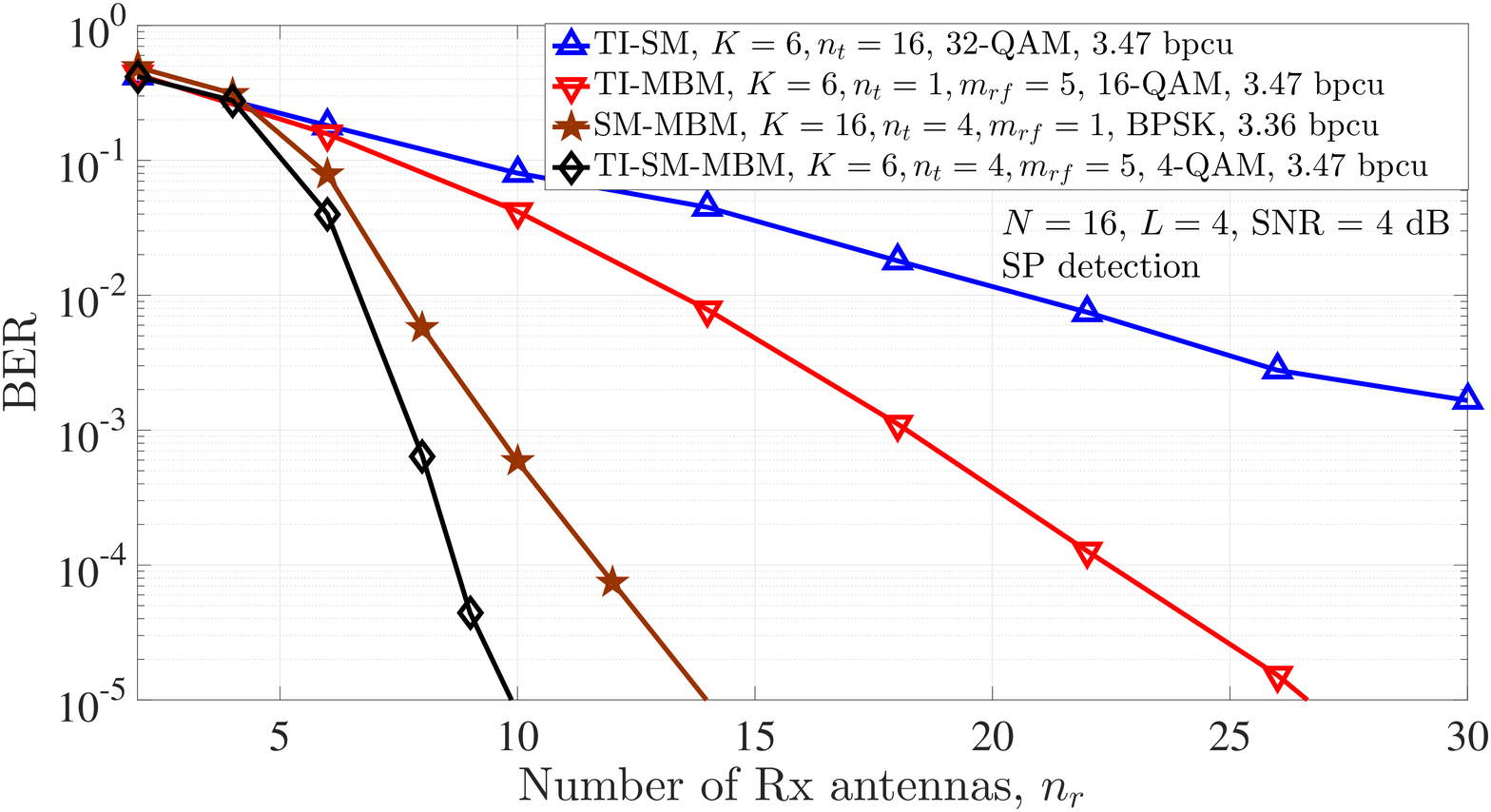}
\vspace{-6mm}
\caption{BER performance of TI-SM, TI-MBM, SM-MBM, and TI-SM-MBM schemes
as a function of $n_r$ with SP based detection at SNR = 4 dB.}
\label{ber_nr}
\vspace{-2mm}
\end{figure}

In Fig. \ref{ber_nr}, we present the BER performance of TI-SM, TI-MBM, 
SM-MBM and TI-SM-MBM schemes as a function of number of receive antennas 
at an SNR of 4 dB. It can be seen  that the BER performance of TI-SM-MBM 
improves drastically with the increase in number of receive antennas 
compared to other schemes. TI-SM-MBM requires $n_r=9$ receive antennas 
to achieve a BER of $10^{-4}$, whereas SM-MBM requires $n_r=12$ and
TI-MBM requires $n_r=22$ receive antennas to achieve the same BER. 
TI-SM exhibits poor performance with an error floor above $10^{-4}$ 
BER. These results illustrate that indexing increased number of transmit 
entities (e.g., as in TI-SM-MBM) can render the possibility of achieving 
a reduction in the required number of receive antennas at the receiver.

\section{Indexed load modulation}
\label{sec4}
In this section, we first introduce the concept of load modulation for 
multiantenna systems \cite{lm1},\cite{lm2}, and then present two indexing 
schemes for load modulated systems, namely, spatially-indexed load 
modulation (SI-LM) and time-indexed load modulation (TI-LM).

\subsection{Load modulation for multiantenna systems}
\label{sec4a}
In conventional RF transmitter hardware, circuit impedance is kept constant 
and a voltage is formed proportional to the information bearing signal. 
This is called the traditional `voltage modulation'. It has the advantage 
that the circuit impedance can be matched to the antenna load impedance 
for optimum power transfer. A disadvantage, however, is that the PA needs 
to be backed-off considerably to ensure a linear response over the entire 
input voltage range. In spatially multiplexed multiantenna systems, separate 
PAs in each RF chain must be driven at high back-offs. RF hardware cost and 
power efficiency concerns thus become impediments to the deployment of 
massive antenna arrays. 
 
\begin{figure}[t]
\centering \includegraphics[width=7cm,height=4cm]{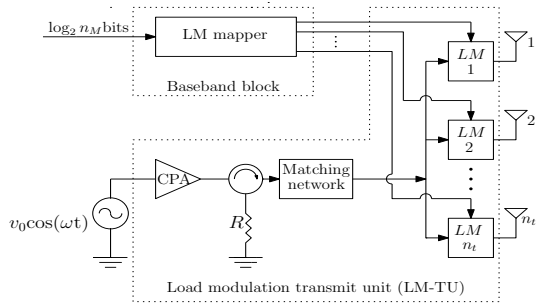}
\vspace{-0mm}
\caption{Load modulation array.}
\label{lmtu_tx}
\vspace{-3mm}
\end{figure}

Load modulation array refers to an array architecture wherein the 
antenna load impedances are chosen to be proportional to the 
information bearing signals while being driven by a sinusoid of fixed 
amplitude and phase and a single central power amplifier (CPA) \cite{lm1}. 
Referring to a multiantenna transmitter, the load impedance in the $l$th 
antenna, $Z_l(t)$, is chosen to be proportional to the $l$th transmit 
signal $s_l(t)$, $l= 1,2,\cdots,n_t$. The circuit that achieves this is 
called a load modulator. A load modulator can be implemented by means 
of varactor 
diodes or pin-diodes. Load modulation thus generates input currents to the 
antennas based on information bearing signals, in effect implementing the 
desired constellation in the analog domain \cite{lm1}. Figure \ref{lmtu_tx} 
shows a load modulation array. 

The effective admittance seen by the power source is the sum of the
admittances of all antenna loads, i.e., 
$Y(t) = \sum_{l=1}^{n_t}\frac{1}{Z_l(t)}$. The single power source 
becomes equivalent to $n_t$ parallel power sources, each with an
average admittance $Y(t)/n_t$. Since $Y(t)$ varies with the 
information bearing signals, there may be a mismatch between
circuit impedance and effective antenna impedance, and so power may 
be reflected back to the CPA. A circulator is therefore used to redirect 
any reflected power to a resistor $R$. For massive antenna arrays (large 
$n_t$), the law of large numbers ensures that the average admittance 
$Y(t)/n_t$ does not vary much even while the individual admittances 
may vary significantly. So the circuit impedance can be matched to the 
average impedance, which results in only a small power being reflected 
back to the CPA. We term the assembly of the CPA, circulator, and $n_t$ 
antennas and their associated load modulators as a {\em load modulation 
transmit unit} (LM-TU) -- see Fig. \ref{lmtu_tx}. 

The efficiency  of the CPA in an LM-TU is determined by the peak to 
average sum power ratio (PASPR), which is the peak to average power 
ratio (PAPR) aggregated over all the antenna elements \cite{lm3}. For 
massive antenna arrays, the PASPR tends asymptotically to one due to 
law of large numbers, and therefore the CPA can be operated at its
highest efficiency. For small number of antennas, however, the PASPR
can be more than one. To obtain a PASPR close to one with small number 
of antennas, it is desired that the sum power radiated by the antennas 
be made constant. A way to achieve this is to use phase modulation on 
the hypersphere (PMH) \cite{lm3}. 

PMH uses points on the $n_t$-dimensional hypersphere to form the LM 
alphabet.  Let $\mathbb{S}_{{\scriptsize{\mbox{lm}}}}$ denote the 
$n_M$-ary LM alphabet, where 
$n_M\triangleq |\mathbb{S}_{{\scriptsize{\mbox{lm}}}}|$. Let 
{\small
${\mathbb S}_{{\scriptsize{\mbox H}}}(n_t,P)=\{{\bf s}\in {\mathbb C}^{n_t} \ | \ \|{\bf s}\|^2=P\}$} 
denote the $n_t$-dimensional complex-valued hypersphere of radius 
$\sqrt{P}$. 
Then, $\mathbb{S}_{{\scriptsize{\mbox{lm}}}}=\{\vs_1,\vs_2,\cdots,\vs_{n_M}\} 
\subset {\mathbb S}_{{\scriptsize{\mbox H}}}(n_t,P)$.
One way to obtain the signal vectors that constitute 
$\mathbb{S}_{{\scriptsize{\mbox{lm}}}}$ is by generating uniformly 
distributed vectors on the hypersphere and clustering them \cite{lm3}.
An $n_t\times 1$ signal vector ${\bf s}$ from 
$\mathbb{S}_{{\scriptsize{\mbox{lm}}}}$ 
chosen based on $\log_2n_M$ information bits gets transmitted in a
channel use by the $n_t$ load modulators as shown in Fig. \ref{lmtu_tx}. 

\begin{figure}[t]
\centering \includegraphics[width=7cm, height=4cm]{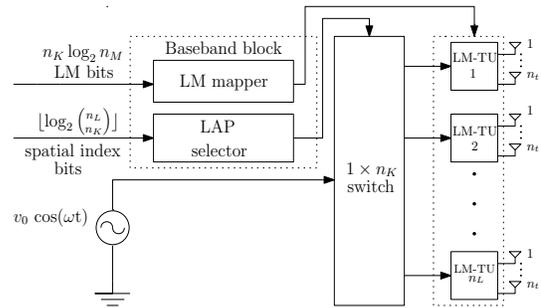}
\vspace{-0mm}
\caption{Spatially-indexed load modulation scheme.}
\label{lmtu_index}
\vspace{-3mm}
\end{figure}

\subsection{Spatially-indexed load modulation (SI-LM)}
\label{sec4b}
It is noted that an LM-TU would require a higher peak-power rated 
CPA as the number of loads driven by it (i.e., $n_t$) is increased. 
For example, it has been shown in \cite{lm2} that an LM-TU with 
$n_t=100$ requires its CPA to have a peak-power rating which is 
about 21 times more than that of a per-stream PA in a conventional 
multiantenna system using spatial multiplexing with $n_t=100$. A higher 
peak-power rated CPA poses challenges in terms of its thermal design 
and number of gain stages \cite{soso}. A tradeoff thereby arises 
between PASPR (which requires a large antenna array) and peak-power 
rating of the CPA (whose design prefers small number of antennas/loads). 
A desirable configuration would then be to have multiple LM-TUs, 
say $n_L$ LM-TUs, each with a CPA that has a lower peak-power 
rating requirement. Transmit signals from all the LM-TUs could 
consequently be spatially multiplexed to obtain the same transmit 
sum power. While this spatial multiplexing of signals from multiple 
LM-TUs is one possibility, another interesting possibility is to 
spatially index the multiple LM-TUs. Such a scheme, which we call
spatially-indexed LM (SI-LM) scheme, is presented in the following.  

Consider a transmitter equipped with $n_L$ LM-TUs, each employing 
$n_t$ antennas (Fig. \ref{lmtu_index}). Every LM-TU is configured to 
generate an $n_M$-ary alphabet on the $n_t$-dimensional hypersphere 
via load modulation. Of the $n_L$ LM-TUs, $n_K$ LM-TUs are activated 
in each channel use. The remaining $n_L-n_K$ LM-TUs remain inactive 
(i.e., OFF). A $1 \times n_K$ switch connects the oscillator output 
to the $n_K$ active LM-TUs. An $n_L$-length pattern of active/inactive 
status of the LM-TUs is called a LM-TU activation pattern (LAP).  
The LAP for a given channel use is chosen based on by 
$\lfloor \log_2{{n_L} \choose {n_K}} \rfloor$ bits. These bits are 
called the spatial-index bits. On each active LM-TU, an LM signal
vector from ${\mathbb S}_{\scriptsize{\mbox{lm}}}$ is transmitted
so that $n_K\log_2n_M$ bits (called the LM bits) are conveyed by the 
$n_K$ active LM-TUs. The achieved rate in the SI-LM scheme therefore 
is given by 
\begin{equation}
\eta_{{\scriptsize{\mbox{si-lm}}}} = \Big \lfloor \log_2{{n_L} \choose {n_K}} \Big \rfloor + n_K\log_2 n_M \quad {\mbox{bpcu}}.
\end{equation}
The SI-LM alphabet, denoted by $\mathbb{S}_{{\scriptsize{\mbox{si-lm}}}}$, 
which is the set of all $n_Ln_t \times 1$-sized vectors that can 
be transmitted is given by
\begin{eqnarray}
{\mathbb S}_{{\scriptsize{\mbox{si-lm}}}} & \hspace{-2mm} = & \hspace{-2mm} \left \{\vs =[\vs_1^T \vs_2^T \cdots \ \vs_{n_L}^T]^T : \vs_j \in \mathbb{S}_{{\scriptsize{\mbox{lm}}}} \cup \mathbf{0}_{n_t}, \right .  \nonumber \\
& \hspace{-2mm} & \|\vs\|_0 = n_Kn_t, 
\ {\mathbf q}^{\mathbf{s}} \in \mathbb{L} \},
\end{eqnarray}
where $\vs_j$ is the $n_t \times 1$ transmitted vector from the $j$th 
LM-TU, $\mathbb{S}_{{\scriptsize{\mbox{lm}}}}$ is the $n_M$-ary LM 
alphabet, ${\bf 0}_{n_t}$ is an $n_t \times 1$ vector of zeros, 
$\|\vs\|_0$ is the $l_0$-norm of $\vs$, $\mathbf{q^s}$ is the LAP 
corresponding to $\mathbf{s}$, and $\mathbb{L}$ is the set of all 
valid LAPs.

Assuming $n_r$ antennas at the receiver, the received signal vector 
$\vy$ can be written as $\vy = \mh\vs + \vn$, where $\mh$ denotes the 
$n_r \times n_Ln_t$ matrix of channel gains such that the gain from 
the $j$th transmit antenna to the $i$th receiver antenna 
$h_{ij} \sim \mathcal{CN}(0,1)$ and $\vn$ is the $n_r \times 1$ noise 
vector with $\mathbf{n} \sim \mathcal{CN}(0,\sigma^2 \mathbf{I})$.

\begin{figure}[t]
\includegraphics[width=9.5cm, height=6cm]{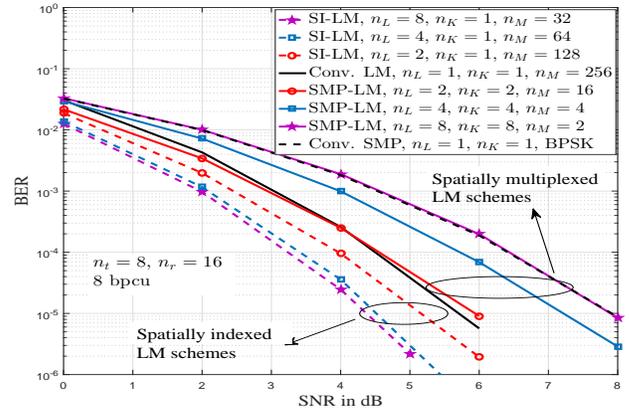}
\vspace{-7mm}
\caption{BER performance of spatially-indexed LM and spatially-multiplexed
LM with $n_t=8$, $n_r=16$, and 8 bpcu in flat fading with ML 
detection. Conventional LM and SMP performance are also shown.}
\label{silm_perf}
\vspace{-4mm}
\end{figure}

{\em ML detection performance:} In Fig. \ref{silm_perf}, we present 
the BER performance of various SI-LM and spatially-multiplexed LM 
(SMP-LM) schemes under ML detection. All schemes are configured to 
achieve 8 bpcu and have $n_r=16$. We also present the BER performance 
of conventional LM without indexing ($n_L = 1$) and conventional SMP
(i.e., SMP without LM) that achieve 8 bpcu. Every LM-TU is assumed to 
have $n_t=8$ antennas (loads). The required LM alphabets are obtained 
on the $n_t$-dimensional hypersphere using the spherical k-means 
clustering (kMC) algorithm \cite{lm3},\cite{kmc}. The following 
observations can be made from Fig. \ref{silm_perf}.
\begin{itemize}
\item 	At $10^{-4}$ BER, we observe that conventional LM (without indexing)
	performs better by about 2 dB compared to conventional SMP with 
	BPSK. By using a signaling alphabet that is simultaneously 
	optimized over $n_t=8$ dimensions, LM is able to achieve this 
	better performance compared to conventional SMP that operates 
	with a per dimension signaling alphabet. 
\item 	When the transmitter is equipped with multiple LM-TUs ($n_L>1$), 
	SI-LM performs better than SMP-LM. For example, SI-LM with 
	$n_L=4, n_K=1$, and $n_M = 64$ performs better by about 2.5 dB
	compared to SMP-LM with $n_L=4, n_K=4$, and $n_M=4$.
	Note that SI-LM achieves this improved performance despite
 	using a larger alphabet size (i.e., $n_M=64$ and 4 for SI-LM and
	SMP-LM, respectively). This is because the effect of spatial 
	interference in SMP-LM outweighs the effect of adding more 
	signal points on the hypersphere in SI-LM. Also, the SNR advantage 
	of SI-LM over SMP-LM increases with $n_L$ since number of index 
	bits increases with increased number of LM-TUs.
\end{itemize}

\subsection{Time-indexed load modulation (TI-LM)}
\label{sec4c}
Motivated by the performance gains offered by time indexing in SM and 
MBM (in Sec. \ref{sec2}), here, we investigate the potential performance 
benefits of time indexing in LM. Towards this, we consider a system model 
which similar to the model considered for TI-SM and TI-MBM in Sec. 
\ref{sec2}. The channel is assumed to be frequency selective with $L$ 
multipaths. Each transmission frame consists of $N+L-1$ time slots, with 
$N$ slots for data transmission and $L-1$ slots for CP. Of the $N$ time 
slots, only $K$ slots are used for data transmission. The remaining 
$N-K$ slots stay silent. The choice of which $K$ slots among the $N$ 
slots are selected for transmission conveys 
$\lfloor \log_2 \binom{N}{K} \rfloor$  information bits. These bits are 
called `time index bits' and the selected time slots are called `active 
slots' in the frame. An $N$-length pattern of active/inactive status of 
the slots in a frame is called a `time-slot activation pattern' (TAP). 
There are $\binom{N}{K}$ possible TAPs, of which 
$2^{\lfloor \log_2 \binom{N}{K} \rfloor}$ are used and they form the 
set of valid TAPs. On each active slot, an LM signal
vector from ${\mathbb S}_{\scriptsize{\mbox{lm}}}$ is transmitted
so that $K\log_2n_M$ bits are conveyed in $K$ active slots. 
The achieved rate in the TI-LM scheme is given by
\begin{equation*}
\eta_{{\scriptsize \mbox{ti-lm}}} = \dfrac{1}{N+L-1} \left [ \Big \lfloor\log _2\binom{N}{K} \Big \rfloor + K\log_2n_M \right ]\quad {\mbox{bpcu}}.
\label{tilm_efficiency}
\end{equation*}
The TI-LM alphabet, denoted by 
${\mathbb S}_{{\scriptsize{\mbox{ti-lm}}}}$, is the set of 
$Nn_t\times 1$-sized vectors obtained by concatenating $N$ vectors each 
of size $n_t\times 1$, as follows:
\begin{eqnarray}
{\mathbb S}_{{\tiny \mbox{ti-lm}}} & \hspace{-2mm} = & \hspace{-2mm} \left \{\mathbf{s} =[\mathbf{s}_1^T \mathbf{s}_2^T\cdots \ \mathbf{s}_N^T]^T : 
\mathbf{s}_j \in \mathbb{S}_{{\scriptsize{ \mbox{lm}}}} \cup \mathbf{0}_{n_t}, \right .  \nonumber \\
&\hspace{-2mm} & \left . \hspace{0mm}  \|\mathbf{s}\|_0 = Kn_t, \ \mathbf{t}^\mathbf{s} \in \mathbb{T}\right \},
\end{eqnarray}
where $\mathbb{T}$ denotes the set of valid TAPs and $\mathbf{t}^\mathbf{s}$ 
denotes the TAP corresponding to $\mathbf{s}$. An $Nn_t \times 1$ 
TI-LM signal vector from $\mathbb{S}_{{\scriptsize{\mbox{ti-lm}}}}$ is 
transmitted over $N$ slots in a frame.

The channel is assumed to remain constant over one frame duration and
to be perfectly known at the receiver. After the removing the CP, the 
$Nn_r \times 1$  received signal vector can be written as 
$\mathbf{y}=\mathbf{Hs}+\mathbf{n}$, where $\mathbf{n}$ is $Nn_r\times 1$  
noise vector with $\mathbf{n}\sim \mathcal{CN}(\mathbf{0},\sigma^2\mathbf{I})$ 
and $\mathbf{H}$ is $Nn_r \times Nn_t$ equivalent block circulant matrix 
having the same form and statistics as in (\ref{block_circulant}).

\begin{figure}[t]
\includegraphics[width= 9.5cm, height=6cm]{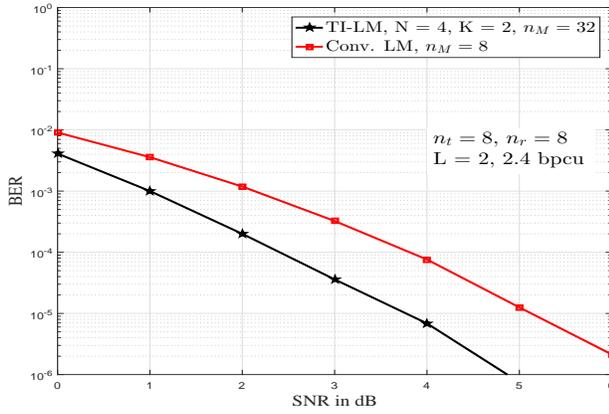}
\vspace{-7mm}
\caption{BER performance of time-indexed LM in frequency selective fading  
with $N=4$, $K=2$, $n_t=8$, $n_M=32$, $L=2$, 2.4 bpcu, $n_r=8$, and ML 
detection. Conventional LM performance is also shown.}
\label{tilm_perf}
\vspace{-5mm}
\end{figure}

{\em ML detection performance:} In Fig. \ref{tilm_perf}, we present 
the BER performance of TI-LM scheme with $N=4$, $K=2$, $n_t=8$, $n_M=32$,
$L=2$, and 2.4 bpcu. The performance of conventional LM (without time 
indexing) with the same 2.4 bpcu is also presented for comparison.
The conventional LM achieves 2.4 bpcu using $N=4$, $n_t=8$, $n_M=8$,
$L=2$. Both systems use $n_r=8$ and ML detection. The LM alphabets are 
obtained from the spherical kMC algorithm as before. At $10^{-4}$ BER, 
it is seen that TI-LM performs better by about 1.5 dB compared to 
conventional LM without time indexing, which is due to reduced ISI
in TI-LM. 

\section{Conclusions}
\label{sec5}
Conveying information bits through indexing transmission entities is an
efficient signaling approach. In this paper, we presented index modulation 
schemes in which transmit antennas, time slots, and RF mirrors are indexed 
simultaneously. ML detection performance of these schemes pointed to very 
good performance possible through indexing of multiple entities. To enable 
the detection of large-dimension signals of these schemes, we exploited 
the inherent sparsity present in their signal vectors. Detection using 
sparsity-exploiting compressive sensing algorithms also showed attractive 
performance. The results indicate that indexing multiple transmission 
entities with a careful choice of system configuration/parameters 
exploiting the underlying tradeoffs involved can be beneficial. We also 
explored the potential benefits of indexing in load modulation. Indexing 
space and time showed promising performance gains in load modulated 
multiantenna systems.


\vspace{-1mm}
\begin{thebibliography}{1}
\vspace{-1mm}
\bibitem{im5g}
E. Basar, ``Index modulation techniques for 5G wireless networks,''
{\em IEEE Commun. Mag.}, vol. 54, no. 7, pp. 168-175, Jul. 2016. 

\bibitem{sm1}
Y. A. Chau and S.-H. Yu, ``Space modulation on wireless fading channels,''
in {\em Proc. IEEE VTC-Fall}, vol. 3, Oct. 2001, pp. 1668-1671.

\bibitem{sm2}
R. Mesleh, H. Haas, S. Sinanovic, C. W. Ahn, and S. Yun, ``Spatial
modulation,'' {\em IEEE Trans. Veh. Tech.}, vol. 57, no. 4, 
pp. 2228-2241, Jul. 2008.

\bibitem{sm3}
M. Di Renzo, H. Haas, A. Ghrayeb, S. Sugiura, and L. Hanzo, ``Spatial
modulation for generalized MIMO: Challenges, opportunities and
implementation,'' {\em Proceedings of the IEEE}, vol. 102, no. 1, 
pp. 56-103, Jan. 2014.

\bibitem{gsm1}
J. Wang, S. Jia, and J. Song, ``Generalised spatial modulation system with
multiple active transmit antennas and low complexity detection scheme,''
{\em IEEE Trans. Wireless Commun}., vol. 11, no. 4, pp. 1605-1615,
Apr. 2012.

\bibitem{gsm2}
T. Datta and A. Chockalingam, ``On generalized spatial modulation,''
in {\em Proc. IEEE WCNC}, Apr. 2013, pp. 2716-2721.

\bibitem{gsm3}
T. Lakshmi Narasimhan and A. Chockalingam, ``On the capacity and 
performance of generalized spatial modulation,'' {\em IEEE Commun. 
Lett.}, vol. 20, no. 2, pp. 252-255, Feb. 2015.

\bibitem{mu_sm1}
N. Serafimovski1, S. Sinanovic, M. Di Renzo, and H. Haas, ``Multiple 
access spatial modulation,'' {\em EURASIP J. Wireless Commun. Netw.}, 
vol. 299, Sep. 2012.

\bibitem{mu_sm2}
T. Lakshmi Narasimhan, P. Raviteja, and A. Chockalingam, ``Large-scale 
multiuser SM-MIMO versus massive MIMO,'' in {\em Proc. ITA}, 
Feb. 2014, pp. 1-9.

\bibitem{mu_sm3}
P. Yang, Y. Xiao, K. V. S. Hari, A. Chockalingam, S. Sugiura, H. Haas, 
M. Di Renzo, Z. Liu, L. Xiao, S. Li, and L. Hanzo, ``Single-carrier spatial 
modulation: a promising design for large-scale broadband antenna systems,''
{\em IEEE Commun. Surveys \& Tutorials}, vol. 18, no. 3, pp. 1687-1716, 
3rd quarter, 2016.

\bibitem{mu_gsm}
T. Lakshmi Narasimhan, P. Raviteja, and A. Chockalingam, ``Generalized 
spatial modulation in large-scale multiuser MIMO systems,'' {\em IEEE 
Trans. Wireless Commun.}, vol. 14, no. 7, pp. 3764-3779, Jul. 2015.

\bibitem{saim1}
N. Ishikawa, R. Rajashekar, S. Sugiura, and L. Hanzo, ``Generalized 
spatial modulation based reduced-RF-Chain millimeter wave communications,''
{\em IEEE Trans. Veh. Tech.}, vol. 66, no. 1, pp. 879-883, Jan. 2017.

\bibitem{saim2}
G. D. Surabhi and A. Chockalingam, ``Efficient signaling schemes for 
mmWave LOS MIMO communication using uniform linear and circular arrays,''
to appear in {\em Proc. IEEE VTC-Spring}, Jun. 2017.

\bibitem{sim1}
R. Abu-alhiga and H. Haas, ``Subcarrier index modulation OFDM,'' in
{\em Proc. IEEE PIMRC}, Sep. 2009, pp. 177-181.

\bibitem{sim2} 
E. Basar, U. Aygolu, E. Panayirci, and H. V. Poor, ``Orthogonal frequency
division multiplexing with index modulation,'' {\em IEEE Trans. Signal 
Process.}, vol. 61, no. 22, pp. 5536-5549, Nov. 2013.

\bibitem{sim3}
Y. Xiao, S. Wang, L. Dan, X. Lei, P. Yang, and W. Xiang, ``OFDM with
interleaved subcarrier-index modulation,'' {\em IEEE Commun. Lett}.,
vol. 8, no. 8, pp. 1447-1450, Aug. 2014.

\bibitem{sim4}
E. Basar, ``On multiple-input multiple-output OFDM with index modulation 
for next generation wireless networks,'' {\em IEEE Trans. Signal Process.},
vol. 64, no. 15, pp. 3868-3878, Aug. 2016.

\bibitem{sim5}
N. Ishikawa, S. Sugiura, and L. Hanzo, ``Subcarrier-index modulation aided 
OFDM - will it work?,'' {\em IEEE Access}, vol. 4, pp. 2580-2593, 2016. 

\bibitem{sim6}
H. Zhang, L-L. Yang, and L. Hanzo, ``Compressed sensing improves the 
performance of subcarrier index-modulation-assisted OFDM,'' {\em IEEE 
Access}, vol. 4, pp. 7859-7873, 2016. 

\bibitem{sim7}
T. Datta, H. Eshwaraiah, and A. Chockalingam, ``Generalized
space-and-frequency index modulation,'' {\em IEEE Trans. Veh. Tech.}, 
vol. 65, no. 7, pp. 4911-4924, Jul. 2016.

\bibitem{tim}
S. Sugiura, S. Chen, and L. Hanzo, ``Generalized space-time shift keying
designed for flexible diversity-, multiplexing- and complexity-tradeoffs,''
{\em IEEE Trans. Wireless Commun.}, vol. 10, no. 4, pp. 1144-1153, 
Apr. 2011.

\bibitem{pim1}
T. Lakshmi Narasimhan, Y. Naresh, T. Datta, and A. Chockalingam,
``Pseudo-random phase precoded spatial modulation and precoder
index modulation,'' in {\em Proc. IEEE GLOBECOM}, Nov. 2014,
pp. 3868-3873.

\bibitem{pim2}
Y. Naresh, T. Lakshmi Narasimhan, and A. Chockalingam, ``Capacity bounds 
and performance of precoder index modulation,'' in {\em Proc. IEEE WCNC}, 
Apr. 2016, pp. 1-6.

\bibitem{am1}
O. N. Alrabadi, A. Kalis, C. B. Papadias, R. Prasad, ``Aerial modulation
for high order PSK transmission schemes,'' in {\em Proc. Wireless 
VITAE}, May 2009, pp. 823-826.

\bibitem{am2}
O. N. Alrabadi, A. Kalis, C. B. Papadias, and R. Prasad, ``A universal
encoding scheme for MIMO transmission using a single active element
for PSK modulation schemes,'' {\em IEEE Trans. Wireless Commun.}, 
vol. 8 , no. 10, pp. 5133-5142, Oct. 2009.

\bibitem{mbm1}
A. K. Khandani, ``Media-based modulation: a new approach to wireless
transmission,'' in {\em Proc. IEEE ISIT}, Jul. 2013, pp. 3050-3054.

\bibitem{mbm2}
A. K. Khandani, ``Media-based modulation: converting static Rayleigh 
fading to AWGN,'' in {\em Proc. IEEE ISIT}, Jun-Jul. 2014,
pp. 1549-1553. 

\bibitem{mbm3}
E. Seifi, M. Atamanesh, and A. K. Khandani, ``Media-based MIMO: 
outperforming known limits in wireless,'' in {\em Proc. IEEE ICC},
May 2016, pp. 1-7. 

\bibitem{mbm4}
Y. Naresh and A. Chockalingam, ``On media-based modulation using RF
mirrors,'' {\em IEEE Trans. Veh. Tech.}, 2016. Available IEEE Xplore: 
DOI: 10.1109/TVT.2016.2620989.

\bibitem{lm1}
M. A. Sedaghat, V. I. Barousis, R. R. Muller, and C. B. Papadias,
``Load modulated arrays: a low-complexity antenna,'' {\em IEEE Commun.
Mag.}, pp. 46-52, Mar. 2016.

\bibitem{lm2}
R. R. Muller, M. A. Sedaghat, and G. Fischer, ``Load modulated 
massive MIMO,'' in {\em Proc. IEEE GlobalSIP}, pp. 622-626,
Dec. 2014.

\bibitem{lm3}
M. A. Sedaghat, R. R. Muller, and C. Rachinger, ``(Continuous) phase 
modulation on the hypersphere,'' {\em IEEE Trans. Wireless Commun.},
vol. 15, no. 8, pp. 5763-5774, Aug. 2016. 

\bibitem{stim}
S. Jacob, T. Lakshmi Narasimhan, and A. Chockalingam, ``Space-time index 
modulation,'' to appear in {\em Proc. IEEE WCNC}, Mar. 2017. 

\bibitem{timbm}
B. Shamasundar, S. Jacob, and A. Chockalingam, ``Time-indexed media-based 
modulation,'' to appear in {\em Proc. IEEE VTC-Spring}, Jun. 2017.

\bibitem{cs1}
J. A. Tropp and A. C. Gilbert, ``Signal recovery from random measurements 
via orthogonal matching pursuit,'' {\em IEEE Trans. Inform. Theory},
vol. 53, vol. 12, pp. 4655-4666, Dec. 2007.

\bibitem{cs2}
D. Needell and J. A. Tropp, ``CoSaMP: iterative signal recovery from 
incomplete and inaccurate samples,'' {\em Applied and Computational
Harmonic Analysis}, 26.3 (2009): 301-321.

\bibitem{cs3}
W. Dai and O. Milenkovic, ``Subspace pursuit for compressive sensing
signal reconstruction,'' {\em IEEE Trans. Inform. Theory}, vol. 55, 
no. 5, pp. 2230-2249, May 2009.

\bibitem{soso}
S. C. Cripps, {\em RF Power Amplifiers for Wireless Communications},
2nd Ed., Artech House, 2006.

\bibitem{kmc}
I.S. Dhillon and D. S. Modha, ``Concept decompositions for large sparse 
text data using clustering,'' {\em Machine Learning}, vol. 42, no. 1, 
pp. 143-175, Jan. 2001. 
\end{thebibliography}
\end{document}